\documentclass[12pt]{article}

\usepackage{graphicx}

\newcommand{\beq}{\begin{equation}}
\newcommand{\eeq}{\end{equation}}
\newcommand{\beqa}{\begin{eqnarray}}
\newcommand{\eeqa}{\end{eqnarray}}
\newcommand{\beqar}{\begin{eqnarray*}}
\newcommand{\eeqar}{\end{eqnarray*}}

\begin{document}
\thispagestyle{empty}

\hfill{\sc UG-FT-246/09}

\vspace*{-2mm}
\hfill{\sc CAFPE-116/09}

\vspace{32pt}
\begin{center}

\textbf{\Large Heavy neutrino decay at SHALON}
\vspace{40pt}

V.G.~Sinitsyna$^{1}$, M.~Masip$^{2}$, S.I.~Nikolsky$^{1}$,
V.Y.~Sinitsyna$^{1}$ \vspace{12pt}

\textit{
$^{1}$P.N.~Lebedev Physical Institute}\\
\textit{Leninsky pr.~53, Moscow, Russia}\\
\vspace{16pt}
\textit{
$^{2}$CAFPE and Departamento de F{\'\i}sica Te\'orica y del
Cosmos}\\
\textit{Universidad de Granada, E-18071 Granada, Spain}\\
\vspace{16pt}
\texttt{sinits@sci.lebedev.ru, masip@ugr.es}
\end{center}

\vspace{30pt}

\date{\today}

\begin{abstract}

The SHALON Cherenkov telescope
has recorded over $2\times 10^6$ extensive air showers during
the past 17 years. The analysis of the signal at different
zenith angles ($\theta$) has included observations from the
sub-horizontal direction $\theta=97^o$. This inclination
defines an {\it Earth skimming} trajectory with 7 km of air
and around 1000 km of rock in front of the telescope.
During a period of 324 hours of observation, after
a cut of shower-like events that may be caused by
chaotic sky flashes or reflections on the snow of vertical
showers, we have detected 5 air showers of TeV energies.
We argue that these events may be caused by the decay
of a long-lived penetrating particle entering the atmosphere
from the ground and decaying in front of the telescope.
We show that this particle can {\it not} be a muon or a tau
lepton. As a possible explanation, we discuss two scenarios
with an unstable neutrino of mass $m\approx 0.5$ GeV and
$c\tau\approx 30$ m. Remarkably, one of these models has
been recently proposed to explain an excess of electron-like
neutrino events at MiniBooNE.

\end{abstract}

\newpage

\noindent
{\bf Introduction}

\noindent Cosmic rays have become a very valuable tool in
astronomy, as they provide a very {\it different} picture of the
sky. In particular, during the past decades gamma-ray
detectors have discovered a large number of astrophysical sources
(quasars, pulsars, blazars) in our Galaxy and beyond. Ground
based telescopes are designed to detect the Cherenkov light of
the shower
produced when a 0.1--100 TeV photon enters the atmosphere. The
light burst in a photon (or electron) air shower has a profile
that can be distinguished from the one from primary protons or
atomic nuclei, which are a diffuse background in such
observations (see \cite{Weekes:2008dr,Voelk:2008fw} for a review).

Cosmic rays may also offer an opportunity to study the
properties of elementary particles. The main objective in
experiments like IceCube \cite{Achterberg:2006md}
or Auger \cite{Abraham:2004dt} is
to determine a flux
of neutrinos or protons as they interact with terrestrial matter.
These interactions involve energies not explored so far at
particle colliders, so their study should lead us
to a better understanding of that physics. In addition,
the {\it size} of the detector and its distance to the
interaction point is much larger there than in colliders,
which may leave some room for unexpected effects
caused by long-lived particles. It could well be that in
the near future cosmic rays play in particle physics a
complementary
role similar to the one played nowadays by cosmology
(in aspects like dark matter, neutrino masses, etc.).

In this paper we describe what we think may be one of such effects.
It occurs studying the response of the SHALON telescope 
\cite{sin_1} to air showers from different zenith angles,
in a sub-horizontal configuration where the signal from
cosmic rays should vanish.

\vspace{0.5cm}
\noindent
{\bf The SHALON mirror telescope}

\noindent SHALON is a gamma-ray telescope \cite{sin_2,sin1} located
at 3338 meters a.s.l. in the Tien-Shan mountain station. It has a
mirror area of 11.2 m$^2$ and a large field of view above $8^o$,
with an image matrix of 144 PMT and a $<0.1^o$ angular resolution.
The recording of Cherenkov light is performed in 50 nsec intervals,
which is enough to acquire complete information about the air shower
while preventing additional light-striking. The trigger is set at
bursts of 8 nsec with a signal in at least 4 PMTs, implying a 0.8
TeV energy threshold on vertical events. The telescope has been
calibrated according to the observation of extensive air showers at
$\theta=0^o$ zenith angle, {\it i.e.}, at an atmospheric depth of 670
g/cm$^2$. Every two-dimensional image of the shower (an elliptic
spot in the light receiver matrix) is characterized by 7 parameters
(widely used in gamma-ray astronomy) defined from the first, second
and third image moments plus the position of the maximum.

SHALON has been operating since 1992 \cite{sin_1,sin1,sin2}, with
the observation of over 2 million extensive air showers. During this
period it has detected gamma-ray signals from well known and also 
from new sources of different type: Crab Nebula, Tycho's SNR,
Geminga, Mkn 421, Mkn 501, NGC 1275, SN2006 gy, 3c454.3 and 1739+522
\cite{sin2,sin3}.

\vspace{0.5cm}
\noindent {\bf Cherenkov bursts below the horizon}

\noindent The study of extensive air showers at large zenith
angles \cite{sin2} has included observations at the sub-horizontal direction
$\theta=97^o$. The configuration of the telescope
is depicted in Fig.~1. The mountain projects a shadow of $\approx
7^o$ over the horizon. The distance of the telescope to the opposite
slope of the gorge is $\approx 7$ km, which corresponds to 16.5
radiation lengths and a depth of 640 g/cm$^2$. From that point the
trajectory finds between 800 and 2000 km of rock before reappearing
in the atmosphere.

\begin{figure}
\begin{center}
\includegraphics[width=9.5cm]{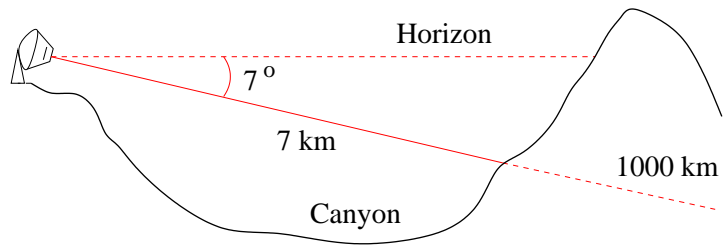}
\end{center}
\caption{Configuration at $\theta=97^o$.}
\label{horizon}
\end{figure}

In Fig.~2 we give the three most recent sub-horizontal events 
recorded by SHALON together with typical vertical air showers 
of similar image parameters.
\begin{figure}[!]
\begin{center}
\begin{tabular}{ccc}
\includegraphics[width=1.7 in]{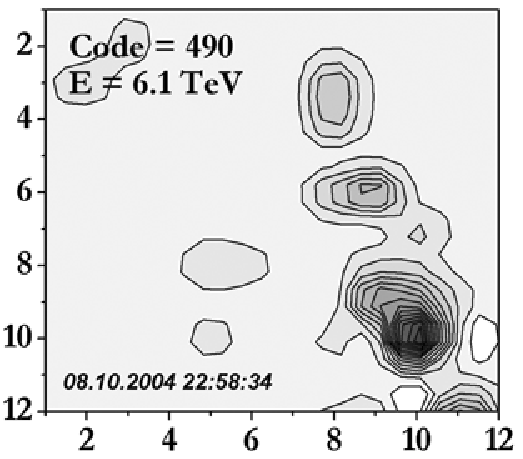}
&
\includegraphics[width=1.7 in]{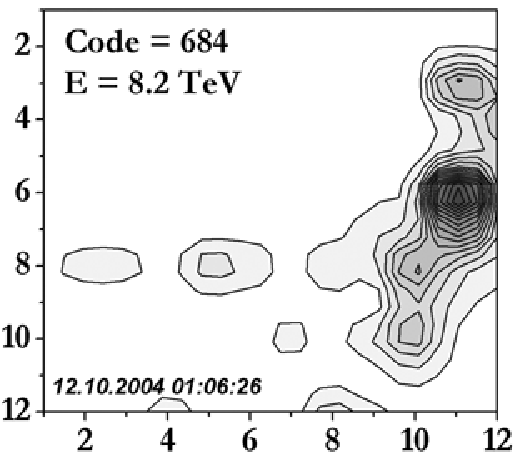}
&
\includegraphics[width=1.7 in]{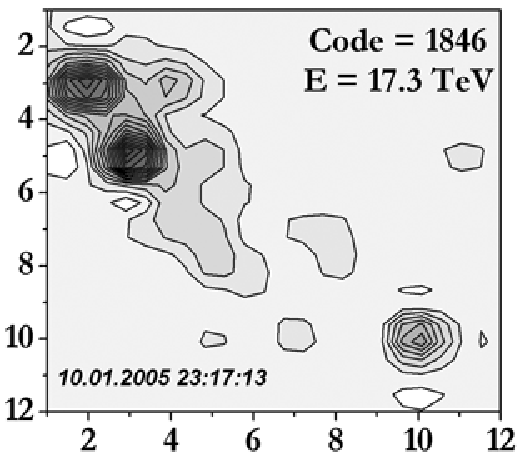}
\\
\includegraphics[width=1.7 in]{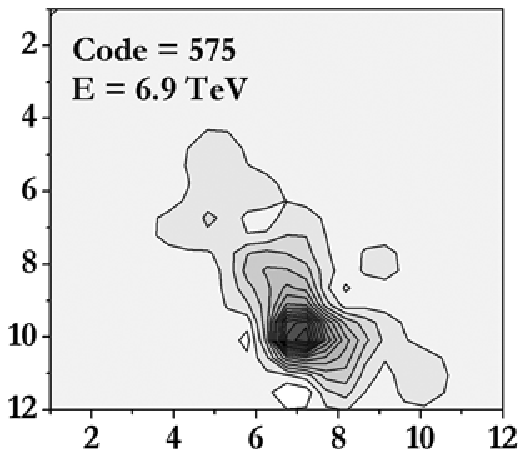}
&
\includegraphics[width=1.7 in]{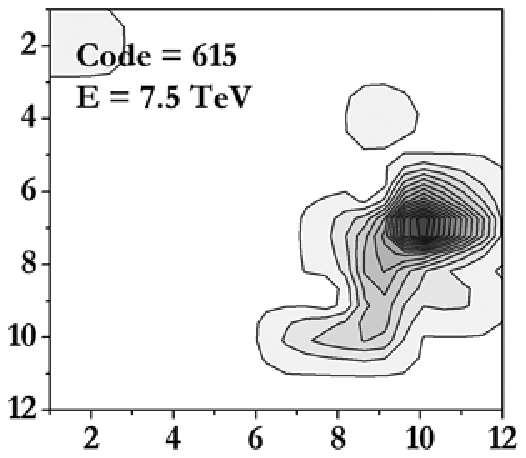}
&
\includegraphics[width=1.7 in]{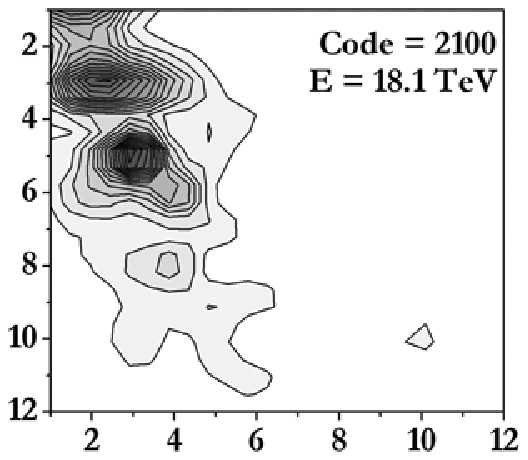}
\end{tabular}
\end{center}
\caption{Events at $\theta=97^o$ (upper)
together with similar air showers from $\theta=0^o$ (lower).}
\label{n_h_spot}
\end{figure}
The grey scale in the plots expresses the number of ADC counts,
whereas CODE is proportional to the shower energy. The five events
look very much like regular extensive showers, they develop within a
narrow angle and are clearly different from the smooth and chaotic
distributions from reflections on the snow or lumniscences of the
atmosphere. In a vertical shower the maximum is $\approx 6$ km above
the telescope, the Moliere radius is $\approx 150$ m, and the
observed angle is $\approx 1.4^o$. If the sub-horizontal showers
start right on the ground their maximum should be at 4 km from the
telescope, defining a Moliere radius of 105 m and an angle of
$1.5^o$. This would make these showers almost indistinguishible from
the vertical ones. Comparing the five events with vertical air 
showers we obtain estimated energies between 6 and 17 TeV (the
values are 11, 7, 6, 8, 17 TeV). 
One should notice, however, that if these events had started 
and developed closer to the telescope their actual energy may
be significantly different.

\vspace{0.5cm}
\noindent {\bf Earth-skimming neutrino interactions}

\noindent The flux of sub-horizontal events is around
$6\times 10^{-6}$
times the flux of TeV cosmic rays reaching the atmosphere.
Such a large flux seems to eliminate the
possibility that these events are due to neutrino interactions in
the air or within the last $\approx 20$ cm of rock. The interaction
length of a 10 TeV neutrino
is $\approx 10^5$ km \cite{Amsler:2008zzb}. This implies that
only one out of $10^{9}$ of them will interact to produce such an
event. The expected neutrino flux from pion and kaon decays at 10
TeV is a per cent fraction of the primary proton flux, whereas the
flux from the prompt decay of charmed hadrons, although uncertain,
should be still smaller at these energies \cite{Costa:2000jw}.
Therefore, the
expected number of events from atmospheric neutrino interactions is
$10^5$ times smaller than the one observed. On the other hand, a
flux of primary (non-atmospheric) neutrinos large enough would
be inconsistent with observations at neutrino telescopes.

Another possibility that can be readily excluded is the decay
in the air of a muon or a tau lepton produced inside the rock.
A 10 TeV muon could emerge if it is produced $\approx 1$ km
inside the rock \cite{Amsler:2008zzb} (one out of $10^5$
incident neutrinos will
produce a muon there). However, the muon decay length at TeV
energies is around $10^4$ km, so the probability that it decays
in the air in front of the telescope is again too small. The
neutrino fluxes required to explain the events from $\mu$ decays
or from $\nu$ interactions are then similar (and excluded).
The probability for tau lepton production in the rock and decay
in the air is not higher. The tau becomes
{\it long-lived} at $\approx 10^8$ GeV. At 10 TeV it should
be produced within the last meter of rock
($c\tau\gamma\approx 0.5$ m), which reduces very much the
number of events.

\vspace{0.5cm}
\noindent {\bf Heavy neutrino decay}

\noindent Therefore, we have to explore possible explanations
based on new physics. The ideal candidate
should be a long-lived massive particle, neutral,
frequently produced in air showers, and very penetrating: able to
cross 1000 km of rock and decay within the 7 km of air in front
of the telescope. If this particle has (possibly suppressed)
couplings to the $W$ and/or $Z$ bosons, its mass $m_h$
should be larger than $m_\mu$ (to decay in the last 7 km of air)
and smaller than $m_\tau$ (to cross 1000 km of rock without
decaying).
Notice that if its decay length at 10 TeV is
$c\tau \gamma \approx 1000$ km, at GeV energies
it will tend to decay far from the detectors in colliders.

An obvious possibility is a sterile neutrino. We take
two Weyl spinors $n$ and $n^c$ and add a Dirac mass term
together with a Yukawa coupling to the lepton family
$L=(\nu_l\; l)$,
\beq
-{\cal L}_\nu= m_n\; n n^c +y_\nu \; h^\dagger L n^c +
{\rm h.c.}
\eeq
Then the Higgs VEV $v$ induces mixing between
$n$ and $\nu_l$:
\beq
-{\cal L}_\nu \supset m_n \;n n^c + m_{EW}
\;\nu n^c = m_h\; \nu_h n^c\;,
\eeq
where $m_{EW}=y_\nu v/ \sqrt{2}$,
$m_h=\sqrt{m_n^2+m_{EW}^2}$,
$\nu_h=c_\alpha n + s_\alpha \nu_l$,
$s_\alpha=m_{EW}/m_h$, and the orthogonal combination
$-s_\alpha n + c_\alpha \nu_l$ remains massless.
The mixing implies couplings of
$\nu_h$ to the $W$ and $Z$ gauge bosons; the
first one will appear suppressed
by $U_{lh}=s_\alpha$, whereas the flavour-changing
(heavy to light)  $Z$ coupling will be
proportional to $c_\alpha s_\alpha$.

A first $\nu_h$ model that we would like to discuss
has been recently proposed by Gninenko
\cite{Gninenko:2009ks} to explain an
anomaly at MiniBooNE \cite{AguilarArevalo:2007it}.
He claims that the excess of
electron-like events in the interactions of the
$\langle E \rangle\approx 800$ MeV $\nu_\mu$ beam
could be caused by the decay of
a heavy neutrino if $m_h\approx 0.5$ GeV,
$c\tau_h \le 30$ m, and $|U_{\mu h}|^2\approx 10^{-3}$.
This explanation requires a large transition magnetic
moment \cite{Mohapatra:1998rq},
$\mu_{tran}\approx 10^{-10}\mu_B$, which implies
a dominant decay mode
\beq
\nu_h\rightarrow \gamma \nu\;.
\eeq
The final photon would convert into a
$e^+e^-$ pair with a small opening angle that would
be indistinguishable from an electron in MiniBooNE.
At the same time, this dominant decay channel could
make the required value of $U_{\mu h}$ consistent with
bounds $|U_{\mu h}|^2\le 10^{-5}$
from BEBC \cite{CooperSarkar:1985nh},
CHARM \cite{Bergsma:1985is} and CHARM2 \cite{Vilain:1994vg},
as these experiments look for decays
into final states with charged particles
($\nu_h\rightarrow e e \nu, \mu e \nu, \mu \pi$).

It is easy to see that such a particle could have an
impact on the SHALON events. At 10 TeV its
decay length is $\lambda_h\approx 600$ km. If $\nu_h$
is produced in the atmosphere with that energy, the
probability that it crosses $\lambda\approx 1000$ km
of rock and decays
within the $\Delta \lambda\approx 7$ km of air in
front of the telescope is
\beq
p=e^{-\lambda/\lambda_h}
\left( 1- e^{-\Delta \lambda/ \lambda_h} \right)
\approx  0.002
\eeq
This implies that the atmospheric flux of heavy neutrinos
should be a 1/1000 fraction of the TeV flux of primary cosmic
rays. This large flux seems difficult to achieve
because $\nu_h$ is not
produced in pion or kaon decays (as $m_h>m_{\pi,K}$), it
appears only in a $|U_{\mu h}|^2\approx 10^{-3}$ fraction of
charmed hadron decays into muons.

A slightly
more frequent production rate could be expected in a second
model, where $\nu_h$
has a sizeable component along the tau neutrino.
NOMAD \cite{Astier:2001ck}
has set limits $|U_{\tau h}|^2\le 10^{-2}$ from
$D_s\rightarrow \tau \nu_h$, and then
$\nu_h\rightarrow \nu_\tau e e$,
but they apply only
to neutrinos lighter than $m_{D_s}-m_\tau\approx 0.19$ GeV.
Cosmological and supernova bounds on $U_{\tau h}$ apply to
lighter values of $m_h$ as well \cite{Dolgov:2000jw}.
On the other hand, LEP bounds
cover just the range $m_h>3$ GeV \cite{Adriani:1992pq}
(decays in the detector of lighter neutrinos are too rare).
Therefore, a possible candidate could have a
0.2--0.4 GeV mass, $|U_{\tau h}|^2\approx 0.1$, and negligible
mixings with the other two families. The dominant decay channels
would be into $\nu_\tau \pi^0$ and into $\nu_\tau e e$,
$\nu_\tau \mu \mu$.
If its decay length at the
TeV energies of the sub-horizontal events is around 1000 km, then
the probability of decay in the air in front of SHALON
is $\approx 0.003$. Its production in air showers would be
through tau decay; one can expect $|U_{\tau h}|^2\approx 0.1$
heavy
neutrinos from each tau produced in the atmosphere. These
tau leptons would mainly come from the prompt decay of
charmed $D_s$ mesons, and also from mesons containing a
bottom quark.
The flux required, a per cent of the TeV proton flux,
seems still too large. Notice, however, that there are
also large uncertainties in the flux and energy of the
sub-horizontal events, or in the tau
production rate in the
atmosphere by cosmic rays \cite{Costa:2000jw}.

\vspace{0.5cm}
\noindent {\bf Summary and discussion}

\noindent
When a cosmic ray enters the atmosphere it produces
an extended air shower with thousands of secondary particles.
Obviously, if there is any new physics it will be
contained in a fraction of these events. Now, if this
{\it exotic} physics includes a long-lived particle, we think
that there is the potential for its discovery in cosmic
ray experiments. Generically, to be detectable
the particle must
{\it survive} after the rest of the shower has been
absorbed by the atmosphere ({\it e.g.}, a long-lived
gluino in horizontal air showers \cite{Illana:2006xg})
or the ground (a stau in neutrino telescopes \cite{Ahlers:2007js}).
In particular, a long-lived neutral particle could
propagate to the center of a neutrino telescope
and start there a contained shower when it decays. However,
this event would look indistinguishible from a standard
neutrino interaction.

In this paper we discuss several air showers obtained at
SHALON in a configuration (see Fig.~1) where the expected number of
events is zero. Around 1000 km of rock absorb the atmospheric flux
of any standard particles but neutrinos. Neutrino interactions in
the rock are frequent, but they are not observable as they {\it
disappear} in just half a meter of soil. A few muons could be
produced during the last km and emerge from the rock, but then the
probability of muon decay within the 7 km of air in front of the
telescope is too small. The crucial difference with a neutrino
telescope is that here the probability of a {\it visible} $\nu$
interaction (in the air or the last centimeters of rock) is
negligible.

We argue that these events may correspond to the decay
of a neutral particle after it is produced in the atmosphere
and has crossed 1000 km of rock. We have studied a couple of
models where this particle is a heavy neutrino,
and have concluded that although 
the required production rate seems
higher than the expected one, due to
a number of uncertainties on the flux and the
energy of the exotic events or on the production of charmed
particles in the atmosphere,
none of these possibilities should be excluded.

\vspace{0.5cm} 
\noindent {\bf Acknowledgements} 

\noindent The 
SHALON Experiment has been supported by RFBR 06-02-17364 
and by the Program of the Russian Academy of Sciences 
"Neutrino Physics" 06, 08, 09. The work of M.M. has
been supported by MEC of Spain (FPA2006-05294) and by Junta de
Andaluc\'\i a (FQM-101 and FQM-437).

\end{document}